\documentclass[aps,twocolumn,showpacs,prl,10pt]{revtex4}
\usepackage{graphicx}
\usepackage{amsmath}
\usepackage{dcolumn}
\usepackage{epsfig}
\usepackage{natbib}
\usepackage{amsfonts}
\usepackage{amssymb}

\begin{document}

\title{The magneto-optical Barnett effect and spin momentum transfer }
\author{A. Rebei}
\email{arebei@mailaps.org}
\author{J. Hohlfeld}
\affiliation{Seagate Research Center, Pittsburgh, PA 15222}
\begin{abstract}
 The interaction of  polarized light with a spin in the 
presence of dissipation is shown to be equivalent to a spin transfer 
process that can cause switching. In plasmas, the spin transfer is 
dominated by a spin-spin exchange term
while at lower energy densities it is dominated by an optical 
Barnett-like effect. This 
latter effect is used in conjunction with optical phonons 
to predict femtosecond magnetization reversal believed to be recently 
measured in GdCoFe thin films. Conventional
approaches based on the  Bloch and the Landau-Lifshitz 
equations do not reproduce this ultrafast switching. 
\end{abstract}
\date{\today}
\maketitle

In 1908, Richardson proposed of what has come to be known as the 
Einstein-de Haas 
effect \cite{richard}. By conservation of angular momentum,
a body which is free to rotate around an axis
 will indeed 
start rotating upon magnetization. A few
years later, the 
inverse effect, that is magnetization by mechanical rotation, was 
proposed and 
observed by Barnett \cite{barnett}. Barnett showed that such 
rotation is 
equivalent to a magnetic field directed along the axis of rotation and 
effectively contributes to the internal field in a magnet. The 
direction of
 magnetization in a body at rest was shown to differ from that in 
a rotating body. A similar effect was also recently observed
in paramagnets subjected to rotating magnetic fields 
instead of mechanical 
rotation \cite{hahn}. Rotating black holes also emit particles (Hawking radiation) 
with net polarization due to the angular momentum of the black hole and can 
also be understood
as a manifestation of a gravitational Barnett effect \cite{hawking}.

  In this paper, we show that there is  an 
optical version of the Barnett effect that can 
lead to magnetization reversal in the 
femtosecond regime. By optical we mean that the laser imparts
spin, rather than orbital momentum, to the 
magnetization. The optical Barnett effect is shown 
to act in similar fashion to a direct spin momentum transfer 
between the laser and the spin polarization of a plasma at high 
power.

We  believe 
that the optical
 Barnett effect has already been observed in a recent 
experiment \cite{rasing}. The 
experiment  demonstrated that it is possible  to switch 
the magnetization of a
 $GdCoFe$ film with a circularly polarized, $40\,$fs, $800 \,$nm
 laser pulse at
 relatively low power, of the order of 
$10^{11}$ W/cm$^2$. Using a single spin picture, 
 we show rigorously 
 that the optical Barnett effect in combination with coupling to
 an appropriate bath, can provide a consistent
explanation of  the observed
femtosecond 
switching result. Since in this case a  real understanding 
of the  short time response of the magnetization 
to the laser is crucial, we refrain from using the
phenomenological  
modified Bloch equations (MB) \cite{hahn} or 
the Landau-Lifshitz (LL) \cite{landau} equation but instead use 
a self consistent formulation for the spin and its 
environment.  

 First we provide a motivation for the 
Hamiltonian that will be adopted for 
the interaction of the laser with the electronic degrees of freedom. 
Starting from the 
Dirac equation, Foldy and Wouthuysen \cite{foldy}
 used a 
series of canonical transformations  to derive
the Pauli equation and other terms of higher order in inverse 
mass. To order $1/m_0^2$, 
the Hamiltonian is given by
\begin{eqnarray}
\mathcal{H} & = & \frac{1}{2m_{0}}\left( 
\mathbf{p}-\frac{e}{c}\mathbf{A}\right)^{2}-e\Phi+\frac{e\hbar}
{2m_{0}c}\mathbf{B}\cdot\sigma \label{ham} \\
& &+
\frac{e\hbar}{8m_{0}^{2}c^{2}} \left\{
\mathbf{p}\cdot\left(  \sigma\times\mathbf{E}\right)  +\left(  \sigma
\times\mathbf{E}\right)  \cdot\mathbf{p}\right\} \nonumber \\
& & +\frac{e^2\hbar}{4m_{0}^{2}
c^{3}}\mathbf{E}\times\mathbf{A}\cdot \sigma
 +\frac{e\hbar}{8m_{0}^{2}c^{2}}\nabla\cdot\mathbf{E}  \nonumber
\end{eqnarray}
This Hamiltonian can also be derived using the gauge-invariant proper-time
method \cite{schwinger}. The terms in this expansion are 
well known. The third and fourth are 
the Zeeman and spin-orbit coupling energies. 
 The fifth, less well known,  term is  
 a Heisenberg exchange-like interaction
between the spin of the 
laser pulse and that of the electron. This 
 term  becomes important at laser powers  of $10^{15}$
 W/cm${^2}$ and above.

  A polarized laser pulse carries angular 
momentum \cite{landau}, but more
importantly it also carries spin angular momentum. The total 
angular momentum
$\mathbf{J}$ of the laser beam occupying volume $V$ is  proportional to 
the momentum
$\mathbf{E}\times\mathbf{B}$ of the wave and given by (cgs)
\begin{equation}
\mathbf{J}=\frac{1}{4\pi c}\int_V d^{3}r\; \mathbf{r\times}\left( 
 \mathbf{E}%
\times\mathbf{B}\right).
\end{equation}
This vector 
can be decomposed into two parts according to  $\mathbf{J}
=\mathbf{L}+\mathbf{S}$, with   $
\mathbf{L}={1}/{4\pi c}\int_V d^{3}r\mathbf{E}\cdot\left(
\mathbf{r}\times\nabla\right)  \mathbf{A}\; $ being the orbital 
contribution,
 while the spin contribution is 
\begin{equation}
\mathbf{S}=\frac{1}{4\pi c}\int_V d^{3}r \; \mathbf{E}\times\mathbf{A}.
\end{equation}
For circularly polarized 
light, with polarization vectors $\mathbf{e}_{\pm}={1}/{\sqrt 2}
(\mathbf{x} \pm i \mathbf{y})$, 
 and a vector potential $
\mathbf{A}\left(  t,\mathbf{r}\right)  =\int {d^{3}k}/{\left( 
 2\pi\right)
^{3}}\left\{  \mathbf{e}_{\mathbf{k}\pm}a_{\pm}
\left(
\mathbf{k}\right)  \exp\left[  i\left(  \mathbf{k}\cdot\mathbf{r}-
\omega
t\right)  \right]  +c.c\right\}$, 
 the spin angular momentum can be written as
\begin{equation}
\mathbf{S}=\frac{1}{2\pi c}\int\frac{d^{3}k}{\left(  
2\pi\right)  ^{3}
}\mathbf{k}\left[  \left|  a_{+}\left(  \mathbf{k}\right)  \right|
^{2}-\left|  a_{-}\left(  \mathbf{k}\right)  \right|  ^{2}\right].
\end{equation}
For a laser with  $
\mathbf{B}_l= \left( B_l  
\cos\omega_{l}t \; , \; B_l \sin\omega_{l}t,\; 0\right) $, 
$a_{-}(\mathbf{k})=0$ and 
 $ a_{+}(\mathbf{k})=0$ correspond to
 positive 
 and negative  
frequencies $\omega_l$
 , respectively. This form clearly shows the chirality of the laser wave 
and most importantly that $\mathbf{S}$ is
parallel to the $\mathbf{k}$-vector which is  
  perpendicular to the 
 $\mathbf{B}$ field. 
 For completeness, we also point out
that for  beams with a finite cross section, the laser
will have an additional component of  $\mathbf{B}$ in the direction
of propagation of the wave which is proportional to the 
gradient of the transverse components \cite{louisell}. In the following
we will ignore this last contribution since it is important 
only at the edges of the beam. Moreover 
the orbital degrees of 
freedom will be considered as part of the environment.

\begin{figure}[th]
\mbox{\epsfig{file=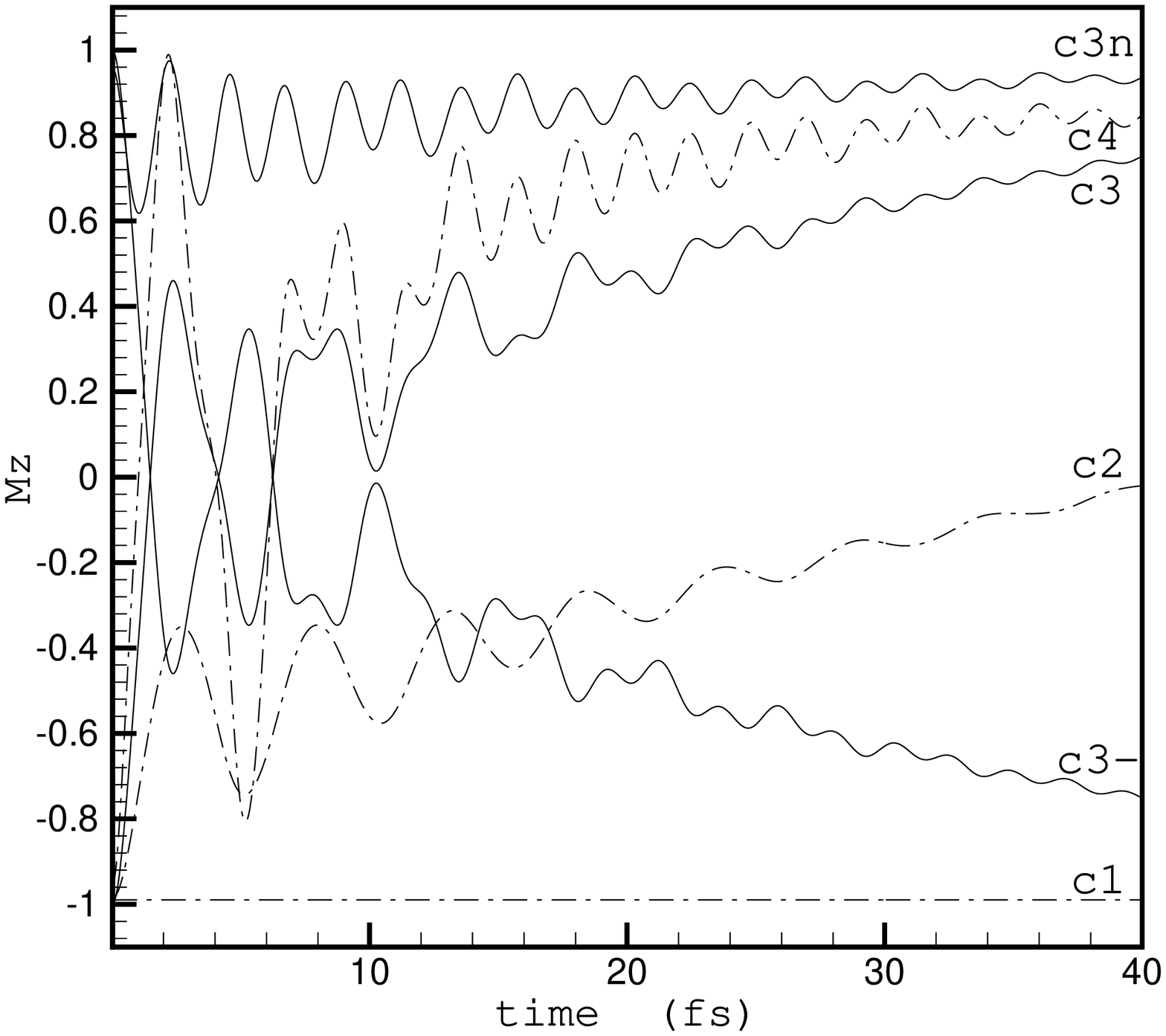,height=8 cm,width=8 cm}}\\[-5.50cm]
\mbox{\hspace*{2.0cm}\epsfig{file=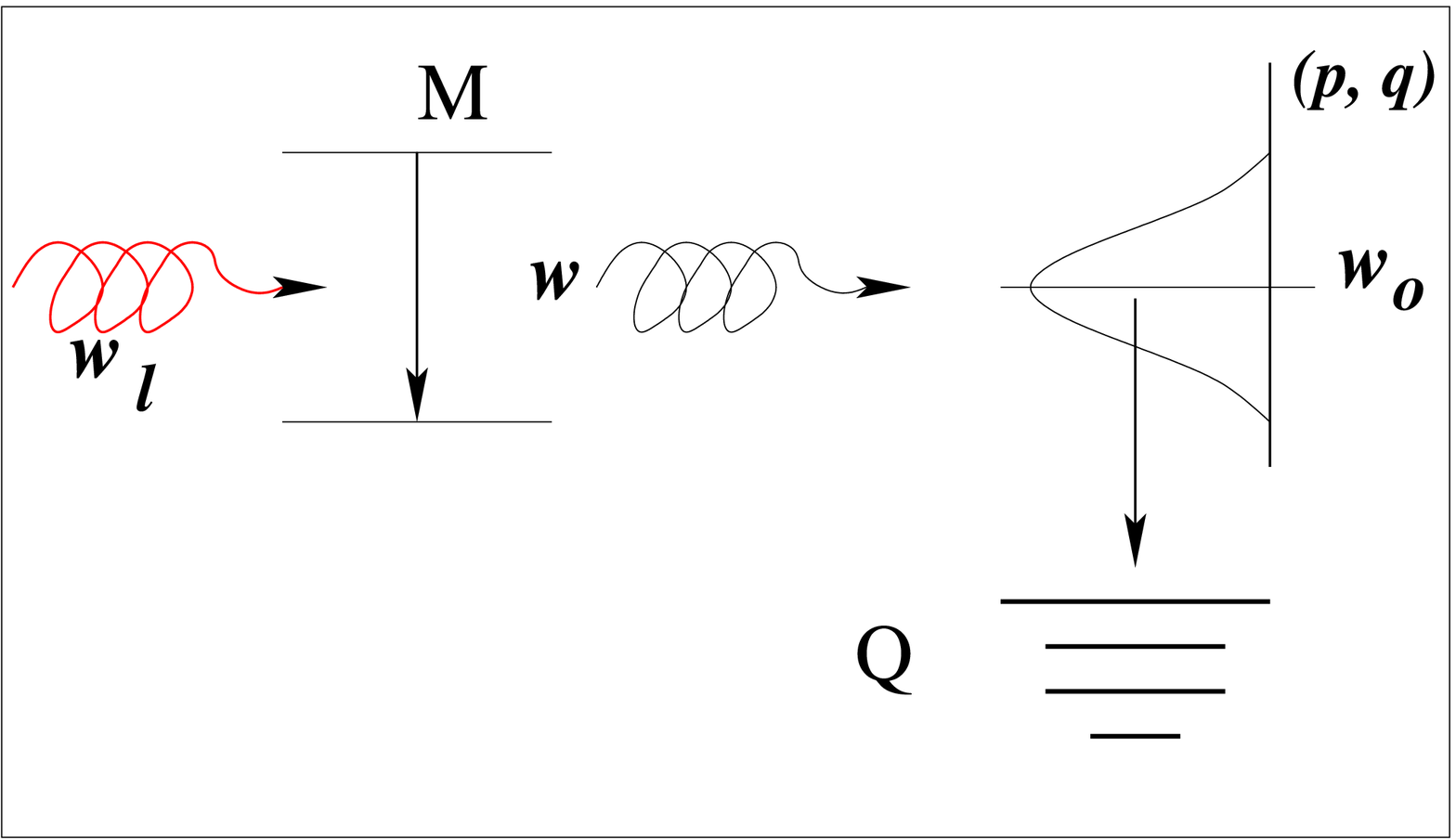,height=3 cm,width=3 cm}}
\vspace{2.5cm}
\caption{ Dynamics of $M_z$, the component of the magnetization 
$\mathbf{M}$ along the Barnett field $\mathbf{H}_B$, induced 
by circularly polarized light with frequency $\omega_l$. For curves $c1$, $c2$, $c3$ and $c4$,
  $\omega_l= 10^{15} Hz$ . For $c3-$,  $\omega_l= -10^{15} Hz$. The laser field  
 $B_l$ is  $10^{-3} H_B$  for $c3$ and $c3-$,  $2 \times 10^{-3} H_B$  for $c4$, $10^{-5} H_B$  for $c1$, and $5 \times 10^{-4} H_B$  for $c2$. $M_z$ 
switches from 
 $-1 (+1)$ to $+1 (-1) $ for positive (negative) 
$\omega_l$ as shown by the curves  $c3$, $c3-$.  The curve $c3n$ shows no switching occurs for  $\omega_l$ and $M_z(0)$ both positive. The different energy transfer channels involved 
in the excitation of
 the magnetization $\mathbf{M}$ by the laser $\omega_l$ and its 
 fast relaxation by the macroscopic bath $\mathbf{Q}$ via an optical
 mode $\mathbf{q}$ with energy $\omega_0 \approx \omega_l$ are
shown in the inset. Presented results are for
 $\omega_0 = 0.8 \omega_l$, $\Gamma=0.2$, and 
$A=10^4 \,$Oe. }
\label{a1}
\end{figure}

In order to give a clear discussion of the optical Barnett effect, 
we introduce an effective Hamiltonian, $\mathcal{H}^{\rm eff}$,  for 
 the spin degrees of freedom derived from
Eq.\,\ref{ham}. $\mathcal{H}^{eff}$  
 includes energy exchange between three different sub-systems
(cf.\,Fig.\,\ref{a1}, inset) and is $(\hbar = 1)$    
\begin{equation}
\mathcal{H}^{eff}=\mathcal{H}_s+\mathcal{H}_q+ \mathcal{H}_{sql}+
\mathcal{H}_Q
\end{equation}
The spin Hamiltonian, $\mathcal{H}_s= -\gamma \mathbf{B}_l \cdot \mathbf{\sigma}  - \frac{1}{2^{3}}A\sigma
_{z}^{2}$, includes the interaction with 
the laser field $\mathbf{B}_l$ and an axial anisotropy 
term. The $\mathbf{\sigma}$'s are Pauli spin matrices for spin $1/2$. The anisotropy term is taken in  
the mean field
approximation to avoid having it trivial for spin 1/2 
fields. It is only  relevant after the 
the laser source is off.  The gyromagnetic ratio $\gamma$ is assumed positive. The 
Hamiltonian, $\mathcal{H}_q=\frac{\mathbf{p}^2}{2} 
+ \frac{\omega_0^2}{2} \mathbf{q}^2 $, represents 
 a single optical phonon mode 
 with energy $\omega_0$ \cite{bloembergen}. The
Hamiltonian 
$\mathcal{H}_Q$ is that of the 
macroscopic bath \cite{schwinger1}. The interaction Hamiltonian 
$\mathcal{H}_{sql}= -\frac{J}{2}
 \mathbf{S}\cdot \mathbf{\sigma}- \lambda(E)
\mathbf{\sigma}\cdot \mathbf{q}  -\mathbf{q}\cdot \mathbf{Q}$, includes an 
exchange term and  a linear coupling to the optical phonon mode. The
 effect of the electric field on the mode $\mathbf{q}$ is implicit in this 
interaction. The
coupling constant in the  
 exchange term is
\begin{eqnarray}
J = \frac{\pi  e^2}{ m^2 c^2}.
\end{eqnarray}
This constant is very
 small
 but since $|\mathbf{S}|$ is proportional to the power, $J S$
 gives rise to fields $\ge 1\,$Tesla  for  
powers $\ge 10^{15}$ W/cm$^2$. While this exchange
 is negligible for the experimental 
conditions in Ref.\,\cite{rasing}, it 
becomes important in plasmas 
\cite{plasma}.  The 
 interaction between
  the spin and the bath is mediated by the 
spin-orbit coupling. The material-dependent 
parameter $\lambda(E)$ is proportional to the power  
 of the laser which 
is  responsible for the excitation of the optical 
modes \cite{bloembergen}. As
first pointed out in Ref.\,\cite{hubner},  a
 strong spin-orbit coupling can
 provide a fast relaxation channel for the spin
of the electrons in the femtosecond regime.  

The coupled 
spin-laser-bath system is better studied in a frame 
$\left(\mathbf{x}_1, 
\mathbf{x}_2,\mathbf{x}_3=\mathbf{z}\right)$ rotating
around the 
$z-$axis 
with  frequency  
 $\omega_l$.  In this frame,  the 
Larmor torque is time-independent and   
the spin Hamiltonian  becomes 
\begin{equation}
\mathcal{H}^{\prime}=-\frac{\gamma B_l}{2}M_{1}  -\lambda \Sigma_i
 M_{i}  b_{i}\left(  t\right) - \frac{\gamma}{2}\left( B_0
 +  H_B \right)M_3
\end{equation}
where $\mathbf{H}_B= \omega_l/\gamma \, \mathbf{z}$ will be 
called the Barnett 
field and      
$B_0= A\left\langle \sigma_{z}\right\rangle + J S_z$. The
 equations of motion for $\langle \mathbf{M} \rangle$, the
average of the 
spin operator, are 
\begin{eqnarray}
\langle \overset{\cdot}{M}_1 \rangle  & =& -\gamma (B_0 +H_B)\langle 
M_2 \rangle + 2\lambda(\langle M_2 b_{3} \rangle - \langle
M_3 b_{2} \rangle ) \label{eq1} \\
\langle \overset{\cdot}{M}_2 \rangle   & =& \gamma (B_0 +H_B) \langle M_1
 \rangle
 - 
\gamma B_l \langle M_3 \rangle + 2 \lambda ( \langle M_3 b_{1} \rangle -
 \langle M_1 b_{3} \rangle ) \nonumber  \\
\langle \overset{\cdot}{M}_3 \rangle  & =& \gamma B_l \langle  M_2 \rangle - 2 
\lambda \left( \langle M_2 b_{1} \rangle -
\langle M_1 b_{2} \rangle \right), \nonumber
\end{eqnarray}
where $b_i$'s are the harmonic mode variables in 
the rotating frame. The equations of motion for $\mathbf{q}$ are
\begin{equation}
(\frac{d^2}{d t^2}+\omega_0^2)\langle q_i \rangle
 = \langle Q_i \rangle + \lambda(E)\langle \sigma_i \rangle \, . \label{eq2}
\end{equation}
The equations for the average $q_i$ are  solved 
 by treating the spin source as a perturbation. The bath $\mathbf{Q}$ is 
assumed Ohmic and is the source of dissipation in the mode $\mathbf{q}$ 
which we take to be $\Gamma=0.2$ using Shen and Bloembergen notation 
\cite{bloembergen}. In the adiabatic limit and in the absence of 
anisotropy, the 
system spin plus bath, $\mathcal{H}_s+
\mathcal{H}_Q $, has been treated earlier, see e.g.\,Ref. \cite{smirnov}. Here we 
 study the more difficult and experimentally relevant 
case of non-adiabatic magnetization reversal.
  
In 
the rotating
 frame, the instantaneous torque is modified by the Barnett 
field. For visible light with  
$\omega_l \approx 10^{15} Hz$, the optical Barnett field is 
as large as $\approx 10^7 \,$Oe. This 
is larger than most exchange 
fields  
and hence a single spin picture should be adequate  even 
for
the treatment of the interaction of a laser with 
a ferromagnet. Despite its tremendous magnitude, the 
Barnett field does not induce femtosecond magnetization 
reversal unless three key requirements are met. First, 
the Barnett field has to be much larger than the laser field 
$\mathbf{B}_l$. Secondly, the magnetization has to be coupled 
to at least one optical mode $\mathbf{q}$ of energy $\omega_0 \approx 
\gamma H_B$. Third, the damping of this mode due to its 
interaction with the macroscopic bath $\mathbf{Q}$ has to 
enable efficient energy transfer from the spins to the 
mode $\mathbf{q}$. All three requirements can be met by various 
combinations of the parameters, $H_B, \, B_l, \, \omega_0$ 
and $\lambda (E)$, but the range of each individual 
parameter depends on the particular choice of the 
others.

 Instead of going to the rotating frame, it is also possible to 
go to a frame which is rotating around the axis of the 
effective field $\mathbf{H}_{\rm eff} = \mathbf{B}_l + (H_B + B_0)\,  
\mathbf{z}$ \cite{abragam}. In this case the transformation 
is explicitly dependent on the Barnett field and is given by 
\begin{eqnarray}
\mathcal{U} = U_{y}\left( \theta \right) U_{z}\left( \omega_s \right) U_{y}^{-1}\left(
\theta \right) U_{z}\left( \omega _{l}\right), \label{fs}
\end{eqnarray}
 with $
U_{\alpha }\left( \epsilon \right)  = \exp \left( -i\epsilon t\sigma
_{\alpha }/2\right) $, $\omega_s=\gamma H_{\rm eff}$ and $
\theta  = \tan ^{-1}\left( \frac{B_{l}}{B_0+H_B }\right)$. It 
can be shown that this transformation $\mathcal{U}$ is 
a spin gauge transformation of the full $U(1)\times SU(2)$ symmetry 
of the Hamiltonian in Eq. \ref{ham} \cite{frohlich}.

 To solve equations \ref{eq1} and \ref{eq2}, we need to calculate the average 
of the product of two operators $\langle \sigma_i(t) q_j(t^\prime)
 \rangle$. For this we need the density matrix of the whole 
system or we may use instead the more powerful functional formalism 
(for a detailed discussion of this method and its application to  
 sd exchange in metals see 
\cite{rebei1} and references therein). The 
generating functional is
\begin{eqnarray}
Z\left[ {\bf J}_{1},{\bf J}_{2}\right]  &=&\left\langle \int D
{\bf \eta }D%
{\bf p}D{\bf q}  \exp \left(   -i\int dt\left( \mathcal{H}_s 
\left( {\bf {\eta},t }\right)
\right.    \right.                    \right.\\ 
& & \left.  \left. \left.
+\mathcal{H}_q \left( {\bf p},{\bf q}\right) +\mathcal{H}_{sql}  -{\bf J}%
_{1}\cdot {\bf q}-{\bf J}_{2}\cdot {\bf \sigma }\right) \right) 
\right\rangle _{Q}
\nonumber
\end{eqnarray}
where the spin variables are written in terms of Grassmann variables,
${\bf \sigma }=-\frac{i}{2}\eta \times \eta $ which allows the use 
of Wick's theorem in the path integral 
expansion and $\mathbf{J}_i$ are 
two virtual external sources. The average values 
are found in the usual way \cite{rebei1,rebei2}
\begin{equation}
\left. \frac{\delta ^{2}\ln Z}{\delta {\bf J}_{1}^{i}\delta {\bf J}_{2}^{j}}%
\right| _{{\bf J}_{1}={\bf J}_2=0}=-\left\langle q^{i}\sigma ^{j}\right\rangle
+\left\langle q^{i}\right\rangle \left\langle \sigma ^{j}\right\rangle .
\end{equation}

 Before switching on the laser, the phonon is 
assumed to be in equilibrium and is not coupled
to the spins. After that, the spin is driven out 
of equilibrium by the laser and $\mathbf{q}$ is treated 
as a perturbation which responds linearly to any changes in the spin. As in the $sd$ exchange problem \cite{rebei1}, the 
equations of motion are non-local 
in time and integrating out the optical mode $\mathbf{q}$ gives 
rise to dissipation and fluctuations even at zero temperature.

Using a Runge-Kutta scheme, Eqs.\,\ref{eq1} and \ref{eq2} are 
solved self-consistently. In line with the discussion above, we find that 
switching occurs only for a limited range of mutually dependent parameters.  The 
results in fig.\,\ref{a1} focus on the power dependence of the magnetic response for 
$\omega_l=10^{15}\,$Hz, $\omega_0=0.8 \omega_l$, $\Gamma=0.2$, and $A=10^4\,$Oe. While
the response to laser fields $B_l \le 10^{-5}H_B$ is negligible on a femtosecond time scale (c1), it 
becomes significant for $5\cdot 10^{-4}H_B$ (c2) and ultrafast reversal is found at higher fields 
of $10^{-3}H_B$ and $2\cdot 10^{-3}H_B$ (c3 and c4, respectively). At the highest fields, 
$M_z$ shows an almost instantaneous reversal accompanied by strong oscillations. The 
period of these oscillations is governed mostly by $H_B$ while their decay depends 
on $\Gamma$. The reversal slows down with time and the approach to equilibrium depends 
on the coupling $\lambda(E)$ of the spins to the mode $\mathbf{q}$ which is proportional 
to the power. The curves c3- and c3n show the strong dependence of the reversal on the chirality 
of the laser.

\begin{figure}[th]
\mbox{{\epsfig{file=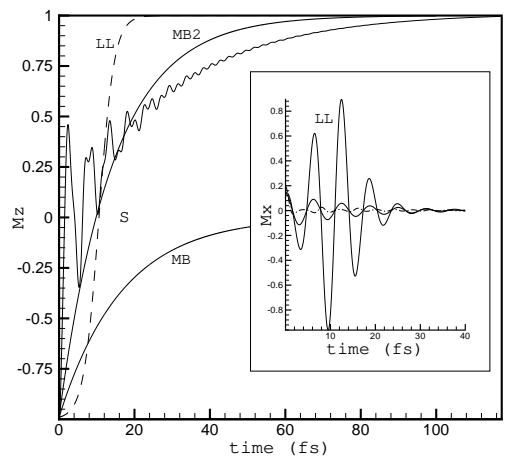,height=7 cm}}}
\caption{ Switching paths for our theory (solid curve), the
corresponding  modified 
Bloch equation MB, MB2 for $1/|\omega_l|T=0.06$,
 and the Landau-Lifshitz (LL) equation for 
$\alpha=0.25$. Zero anisotropy has been assumed for MB, MB2 and 
LL and all other parameters are the same as 
in Fig. \ref{a1}-c3.  Without anisotropy, the 
MB equations lead to the trivial solution $M=0$. Dynamics of
   the transverse 
x-component 
of the magnetization as predicted by our theory (dashed line), 
 MB2, and LL  are 
shown in the inset.  } 
\label{a3}
\end{figure}

 Next we compare our results to those of the 
modified Bloch equations (MB, MB2)
\begin{equation}
\frac{d{\bf M}}{dt}=\gamma{\bf M}\times \left( {\bf H}+{\bf H }_{B}\right) -%
\frac{1}{T}\left( {\bf M}-\chi \left( {\bf H}+  {\bf H }_{B}\right)
\right), \label{mb} 
\end{equation}
and the (modified) Landau-Lifshitz equations (LL)
\begin{equation}
\frac{d{\bf M}}{dt}=\gamma {\bf M}\times \left( {\bf H}+{\mathbf{H} }_{B}\right)
-\alpha \gamma {\bf M}\times \left( {\bf M}\times ({\bf H}+{\bf H}_B )
\right).\label{ll}
\end{equation}
All equations are written in the rotating frame and will 
be solved without taking account of the anisotropy. The anisotropy
is important only after the reversal of the 
magnetization and the laser is off. The common
modified Bloch equations (MB) do
 not include the Barnett field in the relaxation term
 and have been successfully used to interpret low frequency free-induction
decay experiments in paramagnets. However, as shown in Fig. \ref{a3},
MB does not predict switching when the the Barnett field is 
much larger than the laser 
intensity. In contrast, the modified Bloch equations
with the Barnett field accounted for in the relaxation, MB2, 
agree better with our results. 

The LL equations 
with the Barnett field in the relaxation also give rise to {\it fast} 
reversal (Fig. \ref{a3}, discontinuous line). The 
choice of parameters are 
made such that all solutions cross the point  S at the same 
time. In the 
lab frame, the LL 
equations, unlike MB,  
also give reversal but for much longer times. For 
$\alpha =0.25$, a 
switching time of about $40 \, ps$ is found in the lab frame (not shown). It can
 be shown, 
with little effort, that the Larmor equation in the {\it lab frame}
 has a
solution with a non-zero {\it static} z-component such that
  $M_z = \cos\theta$ (cf.  Eq. \ref{fs}).

 The inset in figure \ref{a3} 
shows the relaxation
of the transverse 
components of the magnetization. At short times 
both MB2 and  LL equations fail to capture the real dynamics of 
the reversal process. The failure of LL equations to adequately 
describe the dynamics in our system 
 clearly invalidates the assumption of the universality 
of the LL damping term as was claimed  
 by Koopmans et al \cite{koopman}. Note, that the LL relaxation 
term is also invalid at short times 
when the relaxation is caused by 
momentum relaxation of the conduction 
electrons  \cite{rebei1}.

\begin{figure}[th]
\mbox{\epsfig{file=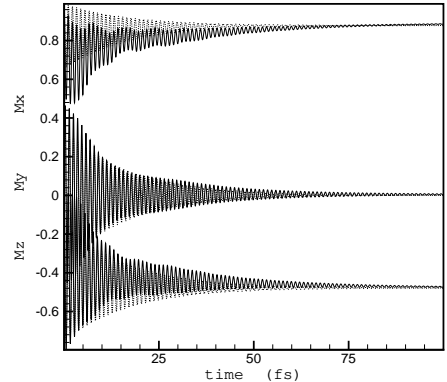,height=6 cm}}
\caption{ Large field solutions for $\omega_l= -10^{15}$ Hz and
  $B_l= 5.0\,H_b$ leading to an exchange field
 of  $1.73 H_B$. Initially the 
$M_y$ (middle) and the $M_z$ (lower)-components are zero. Solutions
of the MB2 equations (dotted lines), with 
$1/|\omega_l|T=0.06$, are plotted on top of our solution.} 
\label{a2}
\end{figure}

Laser powers of the order of  $10^{15} \, W/cm^2$ are relevant only to plasmas since 
most solids are ionized by such intense radiation \cite{price}. In 
this case, we have to abandon the 
identification of the mode $\mathbf{q}$ as that of an optical phonon 
mode and simply assume that it is a term due to collisions between the 
plasma particles. At these intensities,
the spin exchange 
term  $S_z \sigma_z$ starts to become comparable  
to the Barnett field. For
powers in the range 
$10^{17}-10^{20} \, W/cm^2$, full 
reversal of the polarization 
of the magnetized plasma is not possible 
since the 
Zeeman energy term grows much faster than the exchange term. At powers 
higher than $10^{20} \, W/cm^2$, the exchange term dominates all other 
terms and full 
polarization of the plasma along the z-axis becomes again  possible.

 Figure  
 \ref{a2} shows that in the laser configuration studied here $(\mathbf{k} 
\parallel \mathbf{z})$,
the Barnett effect behaves similar to the spin momentum exchange 
term.  The optical Barnett effect is therefore a spin momentum 
transfer 
effect and is not orbital in character 
as in the original experiment of Barnett since it does not 
require charged particles. This can also  be easily 
seen if we write the energy, $\mathcal{H}_l=B_l^2/4\pi$, of the 
laser wave as a 
Zeeman term, 
$\mathcal{H}_l=  1/\gamma V \; \mathbf{H}_B \cdot \mathbf{J}$. The energy 
of the 
laser is not modified in the system studied here. Hence, the laser plays
the role of a second bath that is strongly coupled to the 
spins. The spin momentum is transferred from the laser to the spin through 
the Barnett field as can be seen from the last term in Eq. \ref{ll}.

In summary, we have shown that circularly polarized light
can induce femtosecond magnetization reversal, via spin momentum transfer,  
when optical phonon modes with frequencies 
comparable to those of the light are present. This fast reversal can be recovered 
qualitatively from MB and LL equations only if their relaxation terms are modified to 
account for the Barnett 
field.  The switching is found to be insensitive to the anisotropy which makes the 
proposed mechanism very attractive to high density magnetic recording. In high energy plasmas, 
an additional spin exchange transfer term between the laser 
and the polarization becomes dominant and can be used to 
control the  induced magnetization.


\begin{thebibliography}{99}

\bibitem{richard} O. W. Richardson, Phys. Rev. \textbf{26}, 248 (1908).

\bibitem{barnett} S. J. Barnett, Phys. Rev. \textbf{6}, 240 (1915).


\bibitem{hahn} S-K. Lee, E. L. Hahn, and J. Clarke, Phys. Rev. Lett. 
\textbf{96}, 257601 (2006); M. A. Garstens and J. I. Kaplan, Phys. Rev. \textbf{99},
459 (1955); F. Bloch, Phys. Rev. \textbf{105}, 1206 (1957).

\bibitem{hawking} S. W. Hawking, Commun. Math. Phys. \textbf{43}, 199 (1975). See the discussion 
around Eq. 3.4 where the Barnett field is proportional to $\Omega$, the angular 
velocity of the event horizon, in this case.
\bibitem{rasing} F. Hansteen, C. Stanciu, A. V. Kimel, A. Kirilyuk, and T. Rasing, talk at MMM, Baltimore
 2007; A. V. Kimel et al., Nature \textbf{435}, 655 (2005); F. Hansteen et al., Phys. Rev. Lett. \textbf{95}, 047402 (2005).


\bibitem{landau}L. Landau and Lifshitz, The Classical
Theory of Fields, Pergamon Press 1987.


\bibitem{foldy}L. L. Foldy and S. A. Wouthuysen, Phys. Rev. 
\textbf{78}, 29 (1950). 



\bibitem{schwinger}J. Schwinger, Phys. Rev. \textbf{82}, 664 (1951); A. Rebei 
(unpublished).



\bibitem{louisell} M. Lax, W. H. Louisell, and W. B. McKnight, 
Phys. Rev. A \textbf{11}, 1365 (1975).

\bibitem{bloembergen} Y. R. Shen and N. Bloembergen, Phys. Rev. \textbf{137}, A1787 (1965).

\bibitem{schwinger1}J. Schwinger, J. math. Phys. \textbf{2}, 407 (1961).



\bibitem{plasma} see, e.g.,A. Pukhov, Rep. Prog. Phys. \textbf{66}, 47 (2003).



 

\bibitem{hubner} R. Gomez-Abal, O. Ney, K. Satitkovitchai, 
and W. H\"ubner, Phys. Rev. Lett. \textbf{92}, 227402 (2004). 





\bibitem{smirnov} A. Yu Smirnov, Phys. Rev. B \textbf{60}, 3040 (1999).



\bibitem{abragam} A. Abragam, Principles of Nuclear 
Magnetism, Oxford Press, 1961.

\bibitem{frohlich} J. Frohlich and U. M. Studer, Commun. Math. Phys. 
\textbf{148}, 553 (1992).





\bibitem{rebei1} A. Rebei and M. Simionato, Phys. Rev. B 
\textbf{71}, 174415    (2005).


\bibitem{rebei2} A. Rebei and J. Hohlfeld, Phys. Rev. Lett. 
\textbf{97}, 117601 (2006). 




\bibitem{koopman} B. Koopmans, J. J. M. Ruigrok, F. Dalla Longa, and W. J. M. Longe, 
Phys. Rev. Lett. \textbf{95}, 267207 (2005).

\bibitem{price} D. F. Price {\it et al.\,}, Phys. Rev. Lett. \textbf{75}, 252 (1995).

\end{thebibliography}
\end{document}